**Title:** Title detection: a novel approach to automatically finding retractions and other editorial notices in the scholarly literature

**Short Title:** Automated notice detection in the scholarly literature

## Authors

Ashish Uppala[1], Domenic Rosati[1], Josh M. Nicholson[1], Milo Mordaunt[1], Peter Grabitz[1,2], and Sean C. Rife[1,3]

## Affiliations

1. scite, Brooklyn, NY, USA
2. Charite Universitaetsmedizin Berlin, Berlin, Germany
3. Murray State University, Murray, KY, USA

## Author Note

Address correspondence to: Sean C. Rife, Email: sean@scite.ai

A. Uppala: https://orcid.org/0000-0001-8748-1465
D. Rosati: https://orcid.org/0000-0003-2666-7615

J.M. Nicholson: https://orcid.org/0000-0002-1111-1828

M. Mordaunt: https://orcid.org/0000-0001-5395-4252

P. Grabitz: https://orcid.org/0000-0001-5658-2482

S.C. Rife: https://orcid.org/0000-0002-6748-0841

**Author Contributions**

AU and SCR. wrote code for ingestion of metadata and analyses, and were primarily responsible for writing the manuscript. DR and MM wrote code for ingestion of metadata. PG and JMN provided oversight and review of the research process. All authors reviewed and approved of the submitted draft.

**Funding**

This work was supported by NIDA grant 4R44DA050155-02.

**Competing Interests**

The authors are shareholders and/or consultants or employees of Scite Inc.

**Data Availability**

The datasets, analysis code, and output for the present study are available at https://doi.org/10.5281/zenodo.7094278, https://doi.org/10.5281/zenodo.7093866, and https://doi.org/10.5281/zenodo.7083346.

**Abstract**

Despite being a key element in the process of disseminating scientific knowledge, editorial notices are often obscured and not clearly linked to the papers to which they refer. In the present paper, we describe established methods of aggregating notice data, and introduce a novel method of finding editorial notices in the scientific literature. Specifically, we describe how article titles denote notices to existing publications, and how this information can be used to tie notices to papers in an automated fashion. Finally, as part of a broader movement to make science more transparent, we make notices detected through this and other methods publicly available and describe this dataset and how it can be accessed.

**Introduction**

The process of retracting, withdrawing[1], or appending editorial notices (see Table 1 for definitions of notice types adhered to in the present paper) to published papers is essential to a healthy scientific ecosystem. Unfortunately, this aspect of the scientific process is marked by disorganization and uncertainty. Different publishers and organizations have varying policies regarding when it is appropriate to issue an editorial notice, as well as definitions of the different types of notices [1]. The most consequential type of notice - that of retraction - has become more common in recent years [2], and is often the result of some form of misconduct on the part of the original paper's authors [3], although this may vary from field to field (see also Resnik and Dinse's 2013 work on retractions in response to misconduct [4]).

---

[1] We recognize that not all retractions or withdrawals have a separate editorial notice attached to them [5]. For the sake of simplicity, we simply refer to "editorial notices" as an umbrella term, even in instances when no separate publication can be identified.

Table 1

*Notice types*

| Notice Type | Description |
|---|---|
| Errata | "Published changes or emendations to an earlier article, frequently referred to as corrections or corrigenda, are considered by NLM to be errata, regardless of the nature or origin of the error. Errata identify a correction to a small, isolated portion of an otherwise reliable article. The NLM and other indexing organization do not differentiate between errors that originate in the research process, such as errors in the methodology or analysis, and those that occurred in the publication process, such as typographical mistakes or printing errors. Editors should check with their indexing services for instructions when they have errata related to author names and titles so that online searching issues can be properly addressed." [6] |
| Corrections | Notice issued about an aspect of a published work that is otherwise valid and should therefore remain in the scientific record [6, 7] |
| Expressions of concern | "This indexing term was introduced by the ICMJE and incorporated into the NLM-system in 2004. The expression of concern is a publication notice that is generally made by an editor to draw attention to possible problems, but it does not go so far as to retract or correct an article. An editor who has a significant concern about the reliability of an article but not enough information to warrant a retraction until an institutional investigation is complete will sometimes use an expression of concern." [6] |
| Retractions | "Retraction is a mechanism for correcting the literature and alerting readers to publications that contain such seriously flawed or erroneous data that their findings and conclusions cannot be relied upon. Unreliable data may result from honest error or from research misconduct." [7]<br>"Retractions identify an article that was previously published and is now retracted through a formal issuance from the author, editor, publisher, or other authorized agent. Retractions refer to an article in its entirety that is the result of a pervasive error, nonreproducible research, scientific misconduct, or duplicate publication." [6] |
| Withdrawals | The same as retraction, but (often, but not always) initiated by the author(s) themselves. |

Despite the importance of editorial notices, there is currently no free, open, comprehensive, centralized database of all editorial notices. Indeed, as described below, if one were to rely exclusively on metadata provided by services such as Crossref, a substantial number of notices would be missed. Compounding this problem, publishers are inconsistent in their methods of retracting papers or publishing other types of notices. For example, Lin et al. [8] was retracted three years after publication. In response, The Lancet published a separate retraction notice [9] with "Retraction:" prepended to the title of the original paper. This both updates the public about the revised status of a paper they might have read in the past, and informs those who come across the original paper for the first time that it has been retracted. Publishers may also amend the article landing page and/or PDF to include a watermark indicating that a paper has been retracted (see, e.g. Wakefield et al. [10]). However, many retractions are not so obvious. For example, Tripathi et al. [11] was retracted and a notice of the retraction was published in 2012 [12]. However, neither the article's landing page nor the PDF of the article give any indication that the paper was retracted (see figures 1 and 2). Indeed, the only indications that the paper was retracted were the retraction notice (which is not linked to on the retracted article's landing page) and its entry in PubMed. Essentially, a reader who came across this article outside of PubMed would have no reason to believe it was retracted. One can easily find additional examples of retracted articles of limited visibility [13, 14, 15].

In the present paper, we introduce a set of methods for detecting editorial notices, placing particular emphasis on a novel approach to notice detection that mines metadata and can detect notices that have not been explicitly labeled as such. Our goal was to produce a centralized and open compilation of Digital Object Identifiers (DOIs) to which editorial notices have been attached, the DOI of the notice itself, the type of notice (Expression of Concern, Retraction, etc.), and the date the notice was published.

In our view, data on editorial notices should be openly available, and any editorial notice attached to a given article must be obvious to readers. This should be considered a part of a broader open infrastructure that applies to the entirety of the scientific community and is consistent with larger goals regarding open science and reproducibility [16]. In line with this view, we have made a copy of our notice data, created using the methods outlined below, available to the public at https://doi.org/10.5281/zenodo.7093866. This file will be updated monthly[2].

---

[2] We have also considered making the code we use to generate our records open source, but given the extent to which any attempt to detect notices will be directly tied to a specific infrastructure (type of database, differing programming languages and environments, etc.), we believe that simply describing our methodology and the materials needed to replicate our approach is more appropriate. We have also included pseudocode with additional details in Appendix 1.

Figure 1

*The first page of Tripathi et al. [11], with no indication that the paper has been retracted*

# The role of nitric oxide in inflammatory reactions

Parul Tripathi[1], Prashant Tripathi[2], Luv Kashyap[3] & Vinod Singh[4]

[1]Immunology group, ICGEB, New Delhi, India; [2]Department of Biochemistry, JNU Medical College, Aligarh, Uttar Pradesh, India; [3]Department of Biochemistry, Faculty of Life Sciences, AMU, Aligarh, Uttar Pradesh, India; and [4]Department of Microbiology, Cancer Hospital and Research Institute, Gwalior, Madhya Pradesh, India

**Abstract**

Nitric oxide (NO) was initially described as a physiological mediator of endothelial cell relaxation, an important role in hypotension. NO is an intercellular messenger that has been recognized as one of the most versatile players in the immune system. Cells of the innate immune system – macrophages, neutrophils and natural killer cells – use pattern recognition receptors to recognize the molecular patterns associated with pathogens. Activated macrophages then inhibit pathogen replication by releasing a variety of effector molecules, including NO. In addition to macrophages, a large number of other immune-system cells produce and respond to NO. Thus, NO is important as a toxic defense molecule against infectious organisms. It also regulates the functional activity, growth and death of many immune and inflammatory cell types including macrophages, T lymphocytes, antigen-presenting cells, mast cells, neutrophils and natural killer cells. However, the role of NO in nonspecific and specific immunity *in vivo* and in immunologically mediated diseases and inflammation is poorly understood. This Minireview will discuss the role of NO in immune response and inflammation, and its mechanisms of action in these processes.

## Introduction

The discovery that mammalian cells generate nitric oxide (NO), a gas previously considered to be merely an atmospheric pollutant, is providing important information about many biological processes. In 1992, NO was named the 'molecule of the year,' and various aspects of its biology have since been reviewed extensively (Koshland, 1992; Laskin *et al.*, 1994; Appleton *et al.*, 1996; Christopherson & Bredt, 1997; Bogdan, 1998; Niedbala *et al.*, 1999). With a molecular weight of 30, NO is certainly the smallest molecular mediator (Fang, 1997). When NO formally entered the immunology scene, between 1985 and 1990, its role in the immune system was simply defined as 'Being a product of macrophages activated by cytokines, microbial compounds or both, is derived from the amino acid L-arginine by the enzymatic activity of inducible NO synthase (iNOS or NOS2) and functions as a tumoricidal and antimicrobial molecule *in vitro* and *in vivo*.' Although this basic definition is still accepted, during the past decade it has been recognized that NO plays many more roles in the immune system (Bogdan, 2000). First, in addition to macrophages (Nathan & Hibbs, 1991; MacMicking *et al.*, 1997), a large number of other immune-system cells produce and respond to NO. It exhibits an astonishing range of physiologic functions, from immune defense to blood pressure regulation to the inhibition of platelet aggregation (Lowenstein *et al.*, 1994; Bogdan, 2000). NO is synthesized from the amino acid L-arginine by a family of enzymes, the NOS, through a metabolic route known as the L-arginine–NO pathway (Moncada *et al.*, 1989; Moncada & Higgs, 2002). NO has a short life, between 3 and 20 s in aqueous and oxygen-containing solutions (Moncada *et al.*, 1991) (Fig. 1).

NO has been demonstrated to be a crucial and versatile molecule in the regulation of vascular tone, neurotransmission, acute and chronic inflammation and host defense mechanisms (Michel & Feron, 1997; Maeda & Akaike, 1998; Di Virgilio, 2004). It is involved in innate immunity as a toxic agent towards infectious organisms, but can induce or regulate the death and function of host immune cells, thereby regulating specific immunity (Bogdan *et al.*, 2000a, b). NO may induce toxic reactions against other tissues of the host and because it is generated at high levels in certain types of inflammation (Albina & Reichner, 1995;

Figure 2

*The landing page of Tripathi et al. [11], with no indication that the paper has been retracted*

**About scite**

Scite analyzes the full text of scientific articles, extracts sentences where the authors use their references (citation statements), and uses machine learning to classify the statements as supporting, mentioning, or contrasting the cited claims. The platform also allows users to search citation statements for topics of interest, surfacing key terms in the scientific literature. A key feature of scite is article report pages, which offer a quantitative and qualitative window into how an article has been cited by subsequent papers. Additional information on how scite collects and analyzes data is presented elsewhere [17].

In addition to showing how an article has been cited, scite also displays metadata about a given article. These data are retrieved from Crossref, PubMed, and DataCite on a daily basis. One key piece of information extracted is editorial notices. After experimenting with various partnerships and use of existing sources of metadata, we embarked on building our own method of detecting notices in-house - one that combined these existing sources of data as well as introduced our own method of notice detection - in early 2021.

**Existing Methodology**

A number of tools already provide consolidated lists of publication notices in various formats. For example, RetractionBot [18] uses data from Crossref to identify references to retracted papers in Wikipedia. Open Retractions [19] combines notice data from both Crossref and PubMed and provides an application programming interface (API) for the combined data, while retractcheck [20] is an R package that builds on the Open Retractions API and allows users to automatically check a document for references to retracted works. These are useful tools that provide a valuable service to the scientific community. However, as we discuss below, they are limited in one primary way: they rely on data sources which are, at best, incomplete.

Another, more comprehensive way of aggregating notice data is a journalistic model, best exemplified by the RetractionWatch website and database [21]. At the time of this writing,

RetractionWatch has amassed an impressive database of tens of thousands of notices. Unlike the methods described above, this approach does not rely solely on publishers or editors to modify metadata or publish notices attached to articles. Instead, active curation by the the editors of RetractionWatch and the scientific community supplement traditional sources out new notices through a variety of methods (presumably Internet searches, etc.). This leads to a much more comprehensive collection of publication notices. The downside is that because it relies on manual curation, a journalistic approach is labor intensive and not reproducible (at the time of this writing, their exact methodology has not been made public). We also note that with respect to RetractionWatch in particular, they have explicitly limited the use of their database (see retractionwatch.com/retraction-watch-database-user-guide). We therefore set out to complement earlier methods and work towards an open and comprehensive database that does not rely solely on metadata from publishers and editors, nor requires active, manual curation of publication notices.

**The Scite Approach to Notice Detection**

Scite uses three methods of data collection to create a database of editorial notices. The first two involve aggregating metadata from existing sources, much as the other services discussed above have done. The third is novel, and involves using article titles as indicators of notices. See Figure 3 for a graphical summary of this approach.

First, we leverage metadata from Crossref which indicates whether a given DOI is an update to an existing DOI This update can indicate a newer version of a publication – for example with preprints – or the presence of an editorial notice being attached to the original publication. However, the way these data are provided is somewhat counterintuitive, as it does not involve any straightforward indication that a paper itself has a notice attached to it - rather, it simply indicates that a DOI is a notice in response *to another paper*. In scite's daily metadata updates, we request data through the Crossref API using their from-update-date field. When a notice is published, Crossref includes an update-to record with the update type, date the update was published, and the original DOI to which the update pertains (Figure 4). There is, however, a problem: Crossref relies on publishers or editors taking the initiative to provide and update the relevant metadata (i.e., explicitly indicating that an article is an editorial notice, and providing the DOI of the paper to which the notice refers). Unfortunately, publishers are inconsistent in this practice despite encouragement from Crossref to do so [22]. To further complicate matters, we have found that the update-type field provided to Crossref is sometimes different from the actual notice. As an example, there are notices in Crossref marked as "Addendum" or "Clarification", which in some cases are actually an "Expression of Concern". According to our own internal research, there are 155,336 notices from PubMed that are not included in data from Crossref, 188,966 notices from Crossref that are not included in data from PubMed, and approximately 84,415 notices that are in neither data source (see Figure 5).

*Figure 3*

*The scite.ai notice detection system*

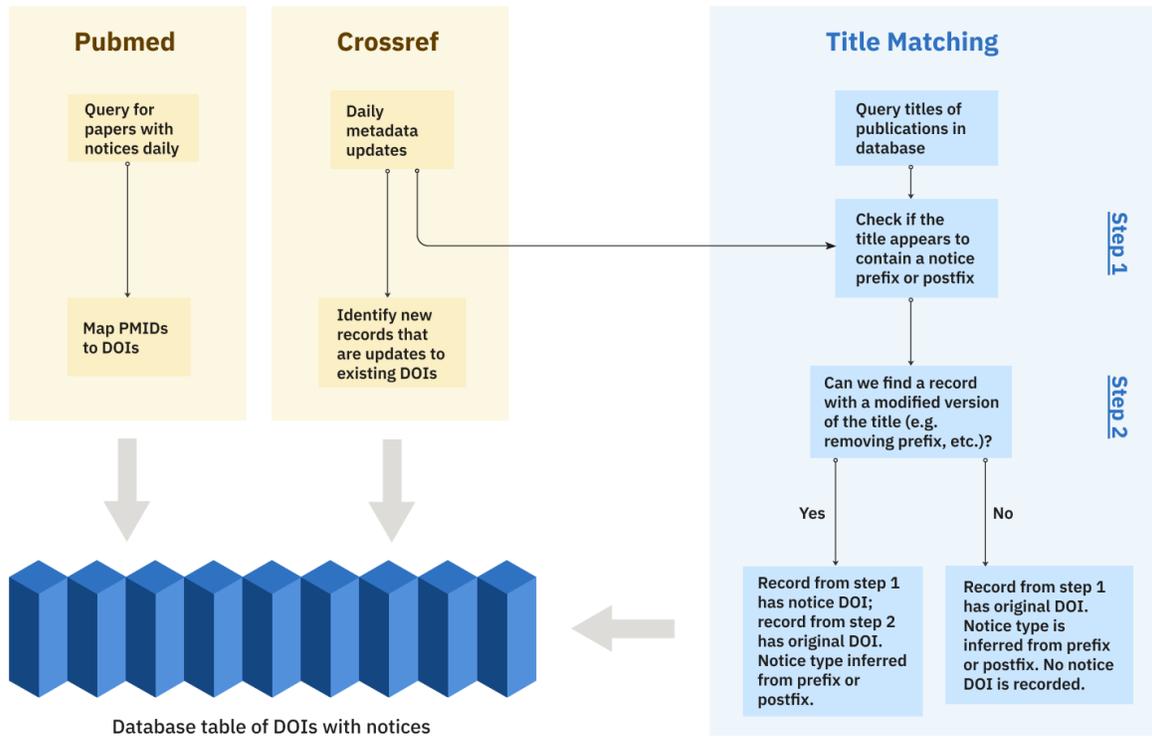

Figure 4

An example payload from Crossref linking two DOIs, one of which is a notice to the original paper. The payload structure usually communicates the notice DOI as an update to an existing paper (as opposed to showing you the notices associated with a particular paper). In some (rare) cases, Crossref might have another record where the DOIs are swapped, creating a circular reference. We deal with this by checking each DOI to determine which one is a notice, and which one is a paper, as described in the Discussion section.

```
{
  "DOI": "10.1111\/wvn.12426",              ← DOI of Notice
  "update-to": [{
    "updated": {
      "date-parts": [
        [2020, 3, 26]
      ],
      "date-time": "2020-03-26T00:00:00Z",
      "timestamp": 1585180800000
    },
    "DOI": "10.1111\/wvn.12418",            ← DOI of Original Paper
    "type": "erratum",
    "label": "Erratum"
  }]
}
```

Figure 5

*Venn diagram representing the number of unique notices found from Crossref, PubMed (via eutils), and the scite Title Detection System.*

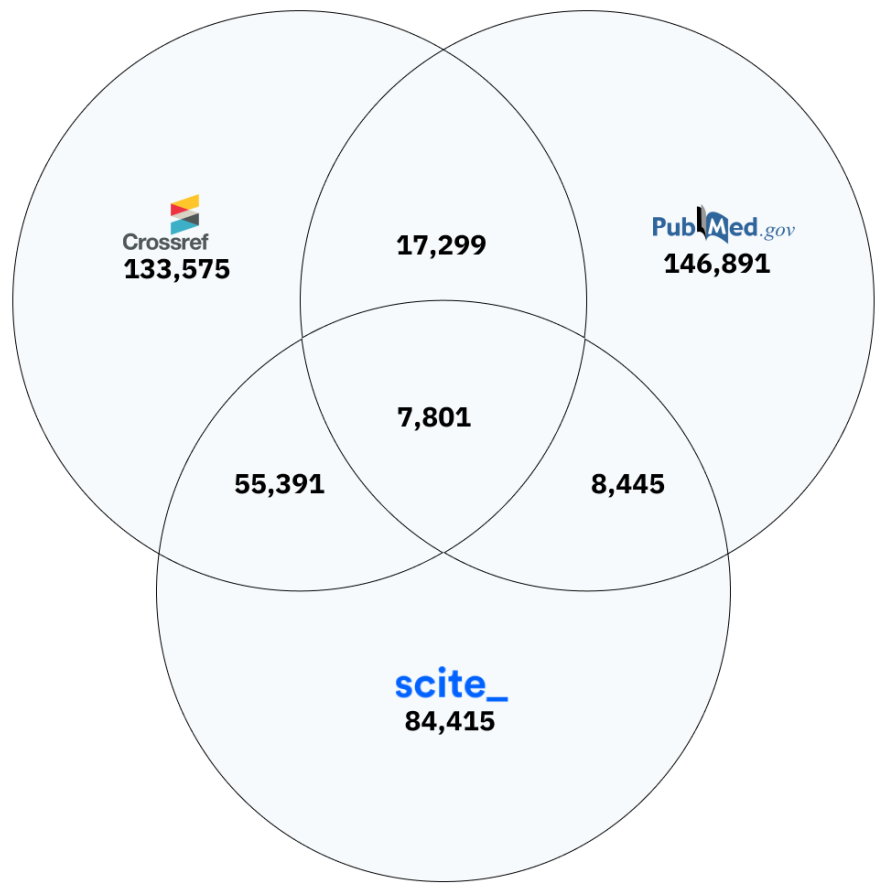

The second source of data is PubMed, which presents notice data in a relatively straightforward manner. The Eutils suite [23] can be used to query the Pubmed database for papers with notices. We use the following search terms, which, in our experience, results in a fairly comprehensive list of notices recorded in PubMed [24]:

- hasexpressionofconcernin
- haserratumin
- hascorrectedrepublishedin
- hasretractedandrepublishedin
- hasretractionin
- retracted+publication[Publication+Type]

The results include PubMed IDs (PubMed's internal identification scheme), which can then be mapped to a DOI using the Pubmed's own API, or by using the downloadable PubMed Central ID (PMC-ID) snapshots that PubMed makes available [25]. Once the DOIs of both the notice and the article to which it refers are in hand, dates can be recorded. Additionally, the data returned by Eutils includes notice type indicators. There are three primary disadvantages of relying on Pubmed for editorial notice data. First, it is limited to a subset of the literature (i.e., biomedical research). Second, like Crossref, PubMed relies on publishers to update their records when a notice is published (A. Sawyer, email communication, January 18, 2022) - although publishers and editors appear to be more consistent in this practice when it comes to PubMed as opposed to Crossref, perhaps because it is a smaller subset of articles and they are more aware of its existence as it is a public-facing tool. Finally, linking PubMed identifiers (PMIDs) to DOIs is a somewhat imperfect process, with errors in publicly-available mappings databases (e.g., https://www.ncbi.nlm.nih.gov/pmc/tools/idconv/; to partially work around this problem, scite extracts mapping data directly using Eutils). As noted above, our data include 188,966 notices from Crossref that do not have a corresponding record from PubMed.

**Title Detection**

The methods listed above are widely known and noted services such as Open Retractions [18] provide this data. However, they are not comprehensive (see Table 3). As such, we began investigating how notices might be detected and attached to publications using existing metadata. One pattern emerged: titles of notices or the articles to which they refer very often reflect a consistent pattern that can be used to detect notices without any explicit indicators. We refer to the use of titles to detect notices as "title detection."

Table 3

*Descriptive data for all notices from various sources. Some sources have types which are unique to them, e.g. Crossref "addendum" and "clarification" which are not present in that form via PubMed or scite. Unfortunately some Addendum or Clarification type notices are actually Expression of Concern, but we expect them to be properly caught via scite's Title Detection.*

| Notice Type | # Total from source | # Unique to source |
|---|---|---|
| *Crossref* | | |
|     Addendum | 1,031 | 1,018 |
|     Clarification | 352 | 352 |
|     Comment | 12 | 12 |
|     Erratum | 53,422 | 24,583 |
|     Correction | 117,529 | 70,421 |
|     Expression of Concern | 901 | 560 |
|     Retraction | 9,125 | 1,009 |
|     Withdrawal | 2,290 | 1,836 |
| *PubMed* | | |
|     Erratum | 135,185 | 108,608 |
|     Comment | 33,027 | 33,027 |
|     Expression of Concern | 1,409 | 1,073 |
|     Retraction/Withdrawal | 10,889 | 4,183 |
| *scite Title Detection* | | |
|     Erratum | 62,134 | 43,248 |
|     Correction | 52,549 | 8,864 |
|     Retraction | 34,801 | 22,701 |
|     Withdrawal | 10,280 | 9,602 |

First, titles of papers that are retracted, withdrawn, corrected, or have errata attached to them are often prepended with a word or phrase to alert readers that there is some issue with that paper. Similarly, notices themselves often begin with a limited set of words or phrases. For example, Albadi et al. [26] published a paper entitled "Click synthesis of 1,2,3-triazole derivatives catalyzed by a CuO–CeO2 nanocomposite in the presence of Amberlite-supported azide" in *Tetrahedron Letters*. Upon being withdrawn, the Crossref record was updated, prepending "WITHDRAWN: " to the title. The DOI landing page for that article was similarly updated, replacing the original article with a withdrawal notice. This change to the paper's metadata was received by Crossref and the "update date" record for that article was altered in the process, thereby making it detectable.

Alternately, some notices are not simple drop-in replacements of the original article - indeed, the majority of notices are published separately. In this case, the notice must be linked to the paper to which it refers, but without relying on official sources of metadata or other groups like Pubmed. This is made possible by searching for alternate versions of the title. For example, Tildesley [27] published "Internet Diabetes Management: A Practical Approach" in *Innovations in Diabetes Care*. In 2016, the paper was withdrawn. In this case, a separate withdrawal notice [28] was published with the title "WITHDRAWN: Internet Diabetes Management: A Practical Approach" and the original article was left online[3]. Searching for papers that begin with "WITHDRAWN: " will not bring up the withdrawn paper, but will return the withdrawal notice.

In order to link the notice to the withdrawn paper, we take the title of the withdrawal notice, remove the prefix "WITHDRAWN: " to get the title of the original paper, and search for it in Crossref to determine its DOI. Alternatively, if a search for the same title without the prepended notice turns up zero results, it may be assumed that the paper with the prepended notice is the withdrawn paper itself (as was the case with the first example).

---

[3] Interestingly, the article's official landing page gives no indication that the paper has been withdrawn - an example of the difficulty in discerning the status of a paper, as discussed above. This is the case for many withdrawals and retractions, although the exact incidence rate is unknown to us at present. One could probably get an estimate of the prevalence of how many retracted articles are not clearly marked as such by scraping and analyzing the text of landing pages for articles known to be retracted.

We have discovered additional patterns of title modifications or notice publications that can be used to find notices and link them to original papers. For example, Wetherby et al. was later retracted, and the notice was titled "Retraction of 'High Activity and Selectivity for Silane Dehydrocoupling by an Iridium Catalyst'" [29]. Thus, in this instance, searching for a title with the text within quotations yielded the original article. A comprehensive list of patterns we have discovered is presented in Table 2. We have made a list of title prepends and postpends we use in our own title detection system available at https://doi.org/10.5281/zenodo.7094278. The same directory also contains a blacklist of titles that our system skips searching for, as they are common but do not contain any information that would identify a particular paper. For example, many errata (over 20,000 at the time of this writing) are issued under the simple title of "Errata", with no other link to the paper in question.

Table 2

*Description of all known notice publication patterns and strategies for linking a notice to the paper to which it refers.*

| Pattern | Primarily Found In | Description | Solution |
|---|---|---|---|
| Simple prepend | Errata, EOC, etc. Limited number of retractions and withdrawals. | Editorial notice is published with an indicator (e.g., "Expression of Concern", "Erratum") prepended to the title of the original article. | Search for a title without the prepend. If found, assume that is the paper to which the notice refers. |
| Alternative title (1) | Retractions, withdrawals | Title is contained in quotations and prepended with "Retraction of" or some similar term. | Search for a title equal to that contained in quotations. If found, assume this is the DOI to which the notice applies. |
| Alternative title (2) | Retractions, withdrawals | Same as Alternative title 1, but abbreviated name of journal is appended to the end. | Same as Alternative title 1. |
| Title append | All notice types | The original title is reprinted (often in quotes) with some indication that it is a notice appended to the end (rather than beginning) of the title. | Same as Alternative title 1, or (if quotations are not present), with the appended notice marker removed. |
| Drop-in replacement | Retractions, withdrawals | Original DOI is replaced with a notice of retraction or withdrawal. No separate notice is published. | If searches for other patterns do not turn up results, assume the given DOI is the retracted article itself. |
| Generic notice | All notice types | A notice is published that has only a simple title (e.g., "Retraction", "Errata") but no indication of what work is being referenced in the title itself. | None yet. May be possible to retrieve additional data from other fields (e.g., Abstract, if available) or the DOI landing page. |

**Discussion**

In the present paper, we have outlined a novel method of detecting editorial notices in the scientific literature. While not entirely comprehensive, this approach allows for a considerable advance in the detection of notices above other methods such as querying an open data source, Crossref or Pubmed, searching for them in a proprietary source such as Scopus or Web Of Science, or merely trusting the notice to be appended on the article of record available online.

There are a number of limitations to our approach. First, we only deal with publication records in the English language. Obviously, it is possible to construct a similar notice-detecting system in other languages, but such an undertaking might be labor intensive depending on how varied notice titles in non-English journals are. Second, our approach will necessarily miss certain notices. This is most apparent in the case of generic notices that do not contain any information in the title that can be used to tie it to another specific paper. For example, 1,360 records in our metadata (primarily from Crossref) are titled "Retraction." These are retraction notices, but their titles do not tie them to any specific paper. At present, we simply record these instances separately for research purposes. In these instances, it may be possible to garner additional information elsewhere in the metadata (e.g., Abstracts). Alternatively, an article's landing page might be scraped in an effort to locate the DOI of the article to which it refers (although doing so might, in some cases, violate a publisher's Terms of Service). Finally, one might look at investigative journalism and social media posts, as publishers may indicate that a paper has been retracted on a blog or social media account.

At the same time, our approach will likely yield a limited number of false positives - that is, notices that are incorrectly tied to an article. This is obviously a serious concern, and we have put in place a number of safeguards in order to ensure that this occurs only rarely. First, a notice must be published after the article to which it refers, and must appear under the same Crossref ID as the article to which it refers. Second, if we find title matches to more than one article, we simply record the multiple matches elsewhere for internal research purposes, but make no further attempts to locate the article to which the notice refers (this is what happens in the generic notices described above). These notices can be searched for and manually resolved later, and the record of the multiple matches can be used to further research and develop automated means of finding notices with multiple title matches. This combination of safeguards has yielded encouraging results: out of a random sample of 100 notices identified through title detection, 98 were correctly tied to the article they reference or the notice itself. The remaining two consisted of retraction notices that were incorrectly identified as the retracted article itself, or incorrect identifications that were due to erroneous/confusing title modifications (e.g., a correction being applied to an erratum rather than the target article itself). Finally, in the rare event that a paper is incorrectly flagged as retracted, we have implemented a whitelist that will prevent our system from identifying specific DOIs as retracted. For example, Ellison et al. [30] is titled, "Retraction of DNA-bound type IV competence pili initiates DNA uptake during natural transformation in Vibrio cholerae," which would normally be flagged as a retracted article, but is skipped because its DOI is whitelisted.

    The largest single category of false positives obtained through title matching is retraction notices being themselves identified as retracted. As noted earlier, we guard against this in part by not searching for title matches when the title contains no useful information (e.g. the title is only "Retraction note", "Notice of retraction", etc.). We also take an additional measure with title matches and query both Crossref and Pubmed to ensure an article marked as having a notice is not, itself, a notice. This combination of automated checks along with occasional manual interventions guards against false positives.

    While we believe our approach is a valuable contribution to science and metascience, we stress that the need for informal methods of notice detection speaks to a larger problem within science as an institution: the stigmatization of retractions, errata, and other types of notices that have negative connotations. At present, publishers, editors, and authors have little

direct incentive to make retractions and errata broadly visible - as indicated by the fact that many retracted papers are not noted as such on their landing pages, and the fact that publishers do not always mark retracted papers as retracted in their article metadata. Rather than being seen as a failure, corrections to the scientific record - whether they be retractions of fatally flawed papers or simple corrections to individual findings - should be regarded as a normal, healthy part of the scientific method. In the clearest case, an author voluntarily withdrawing a paper is a sign of moral integrity and putting the quest for truth above an individual publication record. Thus, the need for creative methods of locating notices is an outgrowth of a deeper problem that is sociological rather than technical.

## References

1. Teixeira da Silva, JA, Dobránszki, J. Notices and policies for retractions, expressions of concern, errata and corrigenda: their importance, content, and context. Science and Engineering Ethics. 2017;23(2):521-554.

2. Steen, RG, Casadevall, A, Fang, FC Why has the number of scientific retractions increased? PloS ONE. 2013;8(7):e68397.

3. Grieneisen, ML, Zhang, M. A comprehensive survey of retracted articles from the scholarly literature. PloS ONE. 2012;7(10):e44118.

4. Resnik, DB, Dinse, GE. Scientific retractions and corrections related to misconduct findings. Journal of Medical Ethics. 2013;39(1):46-50.

5. Drimer-Batca, D, Iaccarino, JM, Fine, A Status of retraction notices for biomedical publications associated with research misconduct. Research Ethics. 2019;15(2):1–5.

6. Counsel of Science Editors. White Paper on Publication Ethics 2021. [Available from: https://www.councilscienceeditors.org/resource-library/editorial-policies/publication-ethics/]. Last accessed 19 September 2022.

7. Wager, E, Barbour, V, Yentis, S, Kleinert, S. Retractions: Guidance from the Committee on Publication Ethics (COPE). Croatian Medical Journal. 2009;50(6):532−535.

8. Lin JC, Lin SC, Mar EC, Pellett PE, Stamey FR, Stewart JA, Spira TJ. Is Kaposi's-sarcoma-associated herpesvirus detectable in semen of HIV-infected homosexual men?. Lancet. 1995;346(8990):1601-1602.

9. Lin JC, Lin SC, Mar EC, Pellett PE, Stamey FR, Stewart JA, Spira TJ. Retraction: Is Kaposi's-sarcoma-associated herpesvirus in semen of HIV-infected homosexual men?. The Lancet. 1998;351(9112):1365.

10. Wakefield AJ, Murch SH, Anthony A, Linnell J, Casson DM, Malik M, Berelowitz M, Dhillon AP, Thomson MA, Harvey P, Valentine A. RETRACTED: Ileal-lymphoid-nodular hyperplasia, non-specific colitis, and pervasive developmental disorder in children. Lancet. 1998;351(9103):637-641

11. Tripathi P, Tripathi P, Kashyap L, Singh V. The role of nitric oxide in inflammatory reactions. FEMS Immunology & Medical Microbiology. 2007;51(3):443-52.

12. Tripathi P, Tripathi P, Kashyap L, Singh V, Bogdan C. The role of nitric oxide in inflammatory reactions (Retraction of vol 51, pg 443, 2007). FEMS Immunology & Medical Microbiology. 2012;66:449.

13. Hayashi S, Lewis P, Pevny L, McMahon AP. RETRACTED: Efficient gene modulation in mouse epiblast using a Sox2Cre transgenic mouse strain. Mechanisms of Development. 2002;1:S97-S101.

14. Pearton SJ, Norton DP, Ip K, Heo YW, Steiner T. Recent progress in processing and properties of ZnO. Superlattices and Microstructures. 2003;34(1-2):3-2.

15. Lotito SB, Frei B. Consumption of flavonoid-rich foods and increased plasma antioxidant capacity in humans: cause, consequence, or epiphenomenon?. Free Radical Biology and Medicine. 2006 Dec 15;41(12):1727-46.

16. Munafò MR, Nosek BA, Bishop DV, Button KS, Chambers CD, Percie du Sert N, Simonsohn U, Wagenmakers EJ, Ware JJ, Ioannidis J. A manifesto for reproducible science. Nature Human Behaviour. 2017;1(1):1-9.

17. Nicholson JM, Mordaunt M, Lopez P, Uppala A, Rosati D, Rodrigues NP, Grabitz P, Rife SC. Scite: A smart citation index that displays the context of citations and classifies their intent using deep learning. Quantitative Science Studies. 2021;2(3):882-898.

18. RetractionBot. [Available from: https://github.com/Samwalton9/RetractionBot]. Last accessed 20 September 2022.

19. Open Retractions. Open retractions: a searchable database of retracted journal articles. [Available from: http://openretractions.com]. Last accessed 19 September 2022.

20. retractcheck. [Available from: https://github.com/libscie/retractcheck]. Last accessed 19 September 2022.
21. Retraction Watch. [Available from: https://retractionwatch.com]. Last accessed 19 September 2022.
22. Meddings, K, Encouraging even greater reporting of corrections and retractions. [Available from: https://www.crossref.org/blog/encouraging-even-greater-reporting-of-corrections-and-retractions]. Last accessed 19 September 2022.
23. Sayers, E. E-utilities quick start. Entrez programming utilities help [Internet]. [Available from: https://www.ncbi.nlm.nih.gov/books/NBK25500]. Last accessed 19 September 2022.
24. PMC User Guide. [Available from: https://www.ncbi.nlm.nih.gov/pmc/about/userguide]. Last accessed 19 September 2022.
25. PubMed User Guide. [Available from: https://pubmed.ncbi.nlm.nih.gov/help]. Last accessed 19 September 2022.
26. Albadi J, Shiran JA, Mansournezhad A. Click synthesis of 1, 4-disubstituted-1, 2, 3-triazoles catalysed by CuO–CeO2 nanocomposite in the presence of amberlite-supported azide. Journal of Chemical Sciences. 2014;126(1):147-150.
27. Tildesley HD, Chow N, White A, Pawlowska M, Ross SA. Internet Diabetes Management: A Practical Approach. Canadian Journal of Diabetes. 2015;39(3):195-199.
28. Tildesley HD, Chow N, White A, Pawlowska M, Ross SA. WITHDRAWN: Internet Diabetes Management: A Practical Approach. Canadian Journal of Diabetes. 2016;40:S1-S13.
29. Wetherby AE, Mucha NT, Waterman R. High activity and selectivity for silane dehydrocoupling by an iridium catalyst. ACS Catalysis. 2012;2(7):1404-1407.

30. Ellison CK, Dalia TN, Vidal Ceballos A, Wang JC, Biais N, Brun YV, Dalia AB. Retraction of DNA-bound type IV competence pili initiates DNA uptake during natural transformation in Vibrio cholerae. Nature Microbiology. 201;3(7):773-780.

## Appendix A: Descriptive approach / pseudo-code for our notice detection system using PubMed, CrossRef, and our own Title Detection System

```
##
## Pseudo-code of our notice detection system
## NOTE: This includes retrieval from CrossRef, PubMed, and our own Title Matching.
##
```

### CrossRef Notice retrieval

We use CrossRef as one of our primary sources of information about papers with a DOI. As part of our daily metadata updates, we query CrossRef using the **from-update-date** and **until-update-date** fields to get a list of DOIs which had any information updated within that window.

An example URL might be:
- https://api.crossref.org/works?filter=from-update-date:2022-01-20,until-update-date:2022-01-22&select=DOI

For each of those DOIs with changes, we then query them using the Works endpoint to get the new metadata.

Most relevant here is the **update-to** field, which is included if a DOI is an update to another paper (i.e. the queried DOI is a notice!).

An example is:
- https://api.crossref.org/works/v1/10.1111/wvn.12426

You will see that the queried DOI (10.1111/wvn.12426) is a notice, and the original paper's DOI is found within the **update-to** field (paper_doi=10.1111/wvn.12418).

That, and information about the notice type is all we need to build a list of information like this, which we then persist into our database.

```
[
    {
    "doi": "<doi_for_publication>",
    "notices": [
        {
        "notice_type": "notice_type",
        "notice_doi": "notice_doi"
        },
        ...
    ]
    },
    ...
]
```

### PubMed Notice retrieval

```
NOTICE_TERMS = [
    'hasexpressionofconcernin',
    'haserratumin',
    'hascorrectedrepublishedin',
    'hasretractedandrepublishedin',
    'hasretractionin',
    'retracted+publication[Publication+Type]'
]
```

We query PubMed's eutils esearch endpoint, in batches of 1000, for each of the notice terms above to get a list of publications (by their PMIDs) having that particular notice.

A sample URL for the eutils esearch might be:
- https://eutils.ncbi.nlm.nih.gov/entrez/eutils/esearch.fcgi?db=pubmed&term={term}&retmode=json&retmax={batch_size}

For each of those PMIDs (which represent publications), we then query PubMed's eutils efetch endpoint to get a list of PMIDs of the associated notices for that publication.

A sample URL for the eutils efetch might be:
- https://eutils.ncbi.nlm.nih.gov/entrez/eutils/efetch.fcgi?db=pubmed&id={pmid}&retmode=xml

Each PMID – of both the publication and its notices – is mapped to a DOI using our internal mappings.

Ultimately, we get a list similar to the one in the CrossRef section above, which we store in our internal database.

### scite Title Notice Detection System

As described in the paper, we also detect whether a paper was retracted or not based on the title, as some publishers prefer to convey notices in this manner.

In our discovery we found various ways that titles can be modified, notably through specific prefixes, or postfixes, on the original title. You can find the current list below, although this may change as publisher behavior changes over time and we find new keywords to include.

```
PREFIXES = [
        THIS ARTICLE HAS BEEN RETRACTED,
        [RETRACTED ARTICLE],
        Retracted:,
        Retracted Article:,
        RETRACTED:,
        Retraction:,
        RETRACTED ARTICLE:,
        WITHDRAWN:,
        Withdrawn:,
        Notice of Retractions:,
        Notice of Retraction:,
        Retraction for:,
        Retraction for,
        Retraction of,
        [RETRACTED],
        [Retracted],
        Erratum to:,
        Erratum:,
        ERRATUM:,
        Retraction notice to,
        Correction to:,
        Correction:,
        Retraction note:,
        Retraction Note:,
        Retracted Article:,
        REMOVED:,
        Removed:,
        Notice of Violation of IEEE Publication Principles:,
        Withdrawal:,
        Retraction notice to [,
        Statement of Retraction:
]

POSTFIXES = [
        [Retracted],
        [RETRACTED],
        [retracted],
        [RETRACTED ARTICLE],
        [Retracted Article],
        [retracted article],
        [Retracted article],
        [THIS ARTICLE HAS BEEN RETRACTED],
```

[This article has been retracted]
]

The general strategy here is to go through the metadata of various DOIs within our system containing a particular prefix or postfix within its title.

For each one, we remove the prefix or postfix to determine the original title, and then try to find the original paper's DOI from that extracted title.

A few scenarios can play out here.

1. We find exactly one result for the extracted title – at which point we treat that matched result as the original publication's DOI, and the DOI we initially found is that of its notice. From the prefix or postfix, we can also infer the type of notice.
2. We find multiple results, in which case we store them for manual review in case any are matches.
3. We find no results, in which case we try an alternate strategy. Specifically, some notices contain the original article within quotes, e.g. **Erratum: "Prevalence, Heritability, and Prospective Risk Factors for Anorexia Nervosa" (2006;63[3]:305-312)**. In this case, in addition to removing the prefix like we did earlier, we also try to search with just the text in the quotes to match on the original title and find its DOI.